\documentstyle[twocolumn,graphics,tighten,prl,aps]{revtex}
\newcommand{\kB}{k}
\newcommand{\kT}{\kB T}

\newcommand{\eqref}[1]{(\ref{#1})}

\newcommand{\myvector}[1]{{\mathbf #1}}
\renewcommand{\r}{\myvector{r}}
\newcommand{\f}{\myvector{f}}
\renewcommand{\v}{\myvector{v}}
\newcommand{\ij}{_{ij}}
\newcommand{\rc}{r_c}
\newcommand{\rd}{r_d}

\newcommand{\TH}{T_H}
\newcommand{\LH}{L_H}
\renewcommand{\Re}{Re}
\newcommand{\rhobar}{\overline\rho}
\newcommand{\figgap}{\vspace{12pt}}
\newcommand{\Beta}{B}

\begin{document}
\draft
\wideabs{
\title{Hydrodynamic bubble coarsening in 
off-critical vapour-liquid phase separation}
\author{Patrick B. Warren}
\address{Unilever Research Port Sunlight, Bebington, Wirral, 
CH63 3JW, UK.}
\date{PREPRINT: 17 April 2001}
\maketitle
\begin{abstract}
Late-stage coarsening in off-critical vapour-liquid phase separation
is re-examined.  In the limit of bubbles of vapour distributed
throughout a continuous liquid phase, it is argued that coarsening
proceeds via inertial hydrodynamic bubble collapse.  This replaces the
Lifshitz-Slyozov-Wagner mechanism seen in binary liquid mixtures. The
arguments are strongly supported by simulations in two dimensions
using a novel single-component soft sphere fluid.
\end{abstract}
} 

Late stage coarsening kinetics in phase separating mixtures shows a
wealth of interesting non-linear coupled phenomena
\cite{Bray,GR,KDBC,WY,Kendon,Tanaka,DPDbin,NC,LJfluid,LSR}.  If
dynamical scaling holds, the growth may be characterised by a scaling
law relating the typical size of domains, $L$, to the elapsed time
$t$, thus $L\sim t^\alpha$.  The standard picture of coarsening in a
space dimension $d=3$ critical quench (equal proportions of the two
phases) is as follows \cite{Bray}. An initial diffusive regime
($\alpha=1/3$) is followed by accelerated growth when hydrodynamic
modes become activated.  If a fluid is characterised by its density
$\rho$, viscosity $\eta$, and surface tension $\sigma$, then
hydrodynamically limited coarsening follows a law like
$L/\LH=f(t/\TH)$ on dimensional grounds, where $\LH=\eta^2/\rho\sigma$
and $\TH=\eta^3/\rho\sigma^2$ are hydrodynamic length and time scales.
If scaling holds, $f(x)\sim x^{\alpha}$.  The expectation is that an
initial viscous hydrodynamic (VH) regime ($\alpha=1$) crosses over to
an inertial hydrodynamic (IH) regime ($\alpha=2/3$), since the
Reynolds number $\Re\equiv\rho L\dot L/\eta=ff'\sim x^{2\alpha-1}$
grows like $x$ in the VH regime and eventually inertial effects must
become important.  The observation that $\Re$ continues to grow in the
IH regime (as $\sim x^{1/3}$) has recently prompted Grant and Elder to
argue that turbulent remixing will limit the asymptotic exponent to
$\alpha\le1/2$ \cite{GR}.  However the currently available numerical
evidence suggests a breakdown of pure scaling is more likely, in the
sense that coarsening is no longer dominated by a single length scale
\cite{KDBC,WY,Kendon}.

In an off-critical quench (unequal proportions of the two phases) the
situation seems much clearer \cite{Bray}, especially if the volume
fraction of the minority phase is vanishingly small \cite{Tanaka}.
For the case of a binary liquid mixture, one rapidly establishes a
population of droplets of the minority phase distributed throughout
the continuous phase.  Such droplets can coarsen via the
Lifshitz-Slyozov-Wagner (LSW) mechanism in which the larger drops grow
at the expense of the smaller drops, as material diffuses through the
continuous phase driven by weak chemical potential
gradients. Diffusion limited droplet coalescence may additionally
occur.  Both mechanisms lead to $\alpha=1/3$ in $d=3$.

Up to now I have not drawn a clear distinction between coarsening in a
single component system quenched into vapour-liquid coexistence, and
demixing in a binary liquid mixture.  Both have similar conserved
order parameters, and one expects similar kinetic coarsening
mechanisms to hold for critical quenches.  In this Letter though, I
argue that coarsening in vapour-liquid systems is fundamentally
different from binary liquid mixtures in the limit of bubbles of
vapour distributed throughout a continuous liquid phase.  This
reflects a fundamental difference in the way the conserved order
parameter is transported, and is motivated by simulations of
coarsening in a novel single-component soft sphere fluid described
below.

Visual observation of the bubble coarsening regime shows that an
LSW-like mechanism operates, but much faster than would be expected
from diffusion of the order parameter.  The large bubbles grow by bulk
movement of fluid, at the expense of the small bubbles which collapse
under the influence of their surface tension.  This is strongly
suggestive of a new hydrodynamically limited bubble coarsening regime,
in which flow fields are generated by pressure gradients caused by a
distribution of bubble sizes.  These pressure gradients are analogous
to the chemical potential gradients in the LSW mechanism, but in this
case the continuous fluid phase responds hydrodynamically.

To gain insight, I consider the paradigmatic problem of an isolated
bubble, initial radius $R_0$, which collapses from rest under the
influence of its surface tension, in zero ambient pressure.  Related
problems of bubble growth and collapse under varying ambient pressure
fields go back to Rayleigh \cite{Rayleigh}, and are of great
technological and military significance in cavitation phenomena
\cite{cavitation}.  In these problems, the flow field is radial,
$\v=(R^2\dot R/r^2)\hat\r$, where $R(t)$ is the drop radius.
Incompressibility requires the $1/r^2$ dependence, and the fluid
velocity is matched to the bubble radius velocity at $r=R$.  This flow
field is \emph{irrotational} and can be derived from a velocity
potential $\phi=-R^2\dot R/r$.  The drop radius obeys the equation
\begin{equation}
\rho(\ddot R+3\dot R^2/2)={p(R)-p(\infty)}
\end{equation}
where $p(r)$ is the pressure.  This can be derived for example by
applying Bernoulli's principle to a streamline extending from the
surface of the drop to infinity \cite{bernoulli}.  In our case
$p(R)=-2\sigma/R$ from the Laplace law (the pressure inside the
droplet is $\approx0$), and $p(\infty)=0$. The above equation can be
integrated to find that the bubble collapses in a time
\begin{equation}
t_{\mathrm{collapse}}=\Bigl(\frac{R_0^3\rho}{\sigma}\Bigr)^{1/2}
\frac{1}{\surd{2}}
\int_0^1\!\frac{dx}{(x^{-3}-x^{-1})^{1/2}}
\end{equation}
(the integral can be evaluated in terms of Euler's $\Beta$-function to
be $\Beta(\frac{5}{4},\frac{1}{2})/2=0.874$).  This result can be
rewritten as $(t_{\mathrm{collapse}}/\TH)^{2/3}\propto R_0/\LH$. It is
not surprising this coincides with the IH scaling law, since we are
only considering a special case of the Navier-Stokes equations in
which viscosity effects are absent.

This result for isolated bubble collapse suggests that in the LSW-like
hydrodynamically limited bubble coarsening regime, the mean bubble
size grows as $\LH({t/T_H})^{2/3}$, identical to the
growth law for IH coarsening.  What is perhaps surprising is the
indication that there is no VH regime, since isolated bubble collapse
is purely inertial (but see discussion below).  The above argument is
for $d=3$, but repeating the analysis in $d=2$ obtains the same
scaling law with additional logarithmic corrections.

I now describe the simulations which motivated the above arguments.
These were undertaken with a novel soft sphere fluid based on the
dissipative particle dynamics (DPD) method \cite{DPD}.  Two features
make DPD attractive for phase ordering studies: firstly one has a
fluid of soft spheres which allows comparatively large time steps to
be taken in the integration algorithm, secondly a momentum-conserving
thermostat is used which preserves hydrodynamic modes.

The soft spheres in DPD interact with short-range forces of the form
$\f\ij=A(1-r\ij/\rc)\hat\r\ij$, acting between all pairs of particles
$i$ and $j$ for which $r\ij<\rc$, where the positions are $\r_i$ and
$\r_j$, $\r\ij=\r_j-\r_i$, $r\ij=|\r\ij|$ and $\hat\r\ij=\r\ij/r\ij$.
Such a force law leads to a predominantly quadratic equation of state
(EOS) which cannot be engineered conveniently to have a van der Waals
loop.  Phase ordering studies, using DPD in its original form, are
therefore limited to binary fluid mixtures where it has been applied
with some success \cite{DPDbin,NC}.

Usually a van der Waals loop results from a hard core repulsion
combined with a long range attraction, for example in previous studies
using Lennard-Jones potentials \cite{LJfluid,LSR}.  However,
introducing hard cores into DPD spoils the attractiveness of the
method.  Therefore an alternative method was pursued which may perhaps
be termed \emph{many-body DPD}.

In many-body DPD, the amplitude of the interaction law is made to
depend on a weighted local density, sampled at the particle positions
in an additional sweep through the pairs of interacting particles.  To
be specific, the following quantity is computed for each particle:
$\rhobar_i=\sum_{j=1}^N w(r\ij)$ ($i=1\ldots N$) where $w(r)$ is a
suitably normalised weight function vanishing for $r>\rc$.  Then the
amplitude of the interaction between a pair of particles $i$ and $j$
is made to depend symmetrically on $\rhobar_i$ and $\rhobar_j$.  In
this way, almost completely arbitrary density dependence can be
introduced into the EOS, including a van der Waals loop.  Many-body
DPD in this form has been described by Pagonabarraga and Frenkel
\cite{PF}, and was also invented independently by Groot \cite{RDG}.
Such density-dependent potentials have featured recently in
coarse-grained models of polymers \cite{BLHM}, and in principle can be
constructed to have arbitrary phase behaviour \cite{ALRT}.

\begin{figure}
\begin{center}
\figgap
\includegraphics{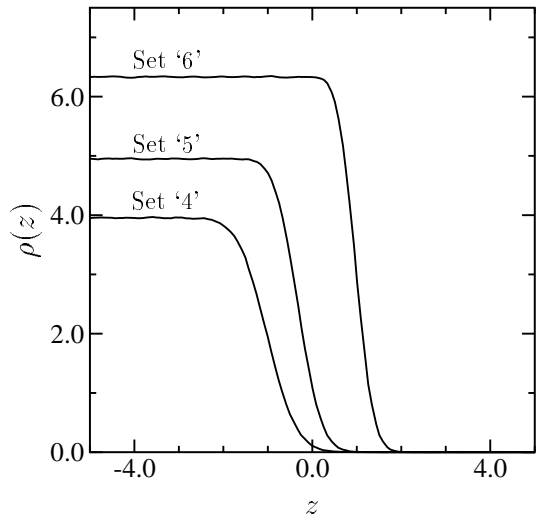}
\end{center}
\caption{Interface density profiles for the three many-body DPD
parameter sets in Table~\ref{table-many}.  The origin of the $z$-axis
is arbitrary.  In all cases there is a fairly sharp interface between
a liquid of moderate density on the left, and an extremely dilute
vapour phase on the right.\label{fig-profiles}}
\end{figure}

The specific model used in the simulations comprises DPD with an
additional interaction force of the form $\f\ij^{\mathrm{extra}} =
B(\rhobar_i+\rhobar_j) (1-r\ij/\rd)\hat\r\ij$ ($r\ij<\rd$).  The $d=2$
weight function is $w(r)=(6/\pi\rd^2)(1-r/\rd)^2$ ($r<\rd$)
\cite{many-note}.  The EOS for this model is then predominantly
\emph{cubic}.  By making $A<0$, $B>0$ and $\rd<\rc$, one can produce a
van der Waals loop and induce vapour-liquid coexistence with a sharp
interface of width $\sim\rc$ (see Fig.~\ref{fig-profiles}).  Three
suitable parameter sets were identified for $d=2$ simulations, and are
given in Table~\ref{table-many} along with the density and viscosity
of the liquid phase, and interfacial tension computed by separate
simulations. In common with previous studies, I fix the units by
setting $m=\rc=\kT=1$ ($m$ is the mass of the particles).

\begin{figure}
\begin{center}
\figgap
\includegraphics{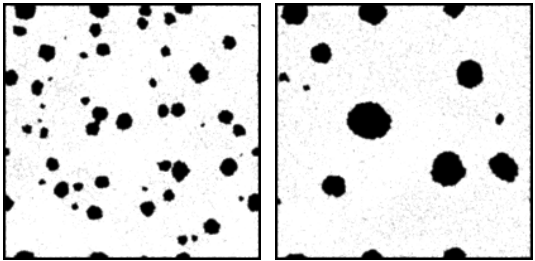}
\end{center}
\caption{Density maps in bubble coarsening regime for set `6' in
Table~\ref{table-many}, at reduced times $t/\TH\approx30$ (left) and
120 (right). Grey scale runs linearly between $\rho=0$ (black) and
$\rho\ge6$ (white).\label{fig-images}}
\end{figure}

Armed with this model, $d=2$ simulations of vapour-liquid phase
ordering were undertaken by preparing a random distribution of
particles in a square domain at a density $\phi\rho$ where $\phi$ is
the desired liquid phase area fraction.  Typical simulations comprised
5--10 runs of $10^4$--$10^5$ particles in a domain of side $100\rc$,
at $\phi=0.9$.  These were run out to times of order 150 (DPD units)
where only one or two large vapour bubbles remain.  The progress of
the simulation was monitored by overlaying a grid at a resolution
$0.5\rc$, computing the local density in each grid element (pixel),
and determining connected sets of pixels where the density is less
than a critical value (usually $\rho=1$) by cluster analysis (see
Fig.~\ref{fig-images}).  These clusters are identified with bubbles,
and statistics on the bubble size distribution were obtained. Most
attention was paid to how the mean bubble area $\langle A\rangle$
grows with time; very similar results were obtained with $\langle
A^{1/2}\rangle$.

\begin{figure}
\begin{center}
\figgap
\includegraphics{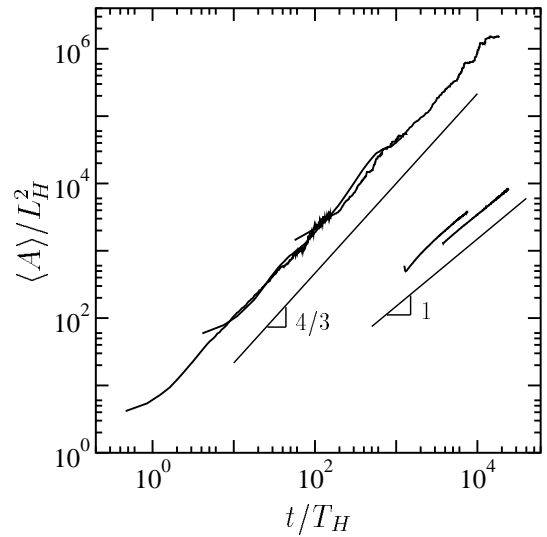}
\end{center}
\caption{Scaling collapse of mean bubble area $\langle A\rangle$ as a
function of time, non-dimensionised by the hydrodynamic length and
time scales, for the three many-body DPD parameter sets in
Table~\ref{table-many} (upper data curves).  Also shown is the
corresponding reduced data for droplet area growth in binary liquid
demixing, for the two conventional DPD parameter sets in
Table~\ref{table-binary} (lower data curves; the lowest being set
`4b').\label{fig-scaling}}
\end{figure}

Fig.~\ref{fig-scaling} shows data from runs for all three parameter
sets, reduced using the hydrodynamic length and time scales.  There
are two key points to note.  Firstly the data collapses onto a
\emph{single curve} despite the order of magnitude variation in $\LH$
and $\TH$ across the sets.  This is a strong indication that bubble
coarsening is hydrodynamically limited.  Secondly the data follows a
scaling law virtually indistinguishable from that expected in an IH
regime \emph{over nearly four decades in reduced time}.  This is a
strong indication that the above arguments concerning
inertia-dominated bubble collapse are correct.  To be precise, the
scaling law found by fitting across all data sets is $\langle A\rangle
= (4.2\pm0.5)\LH^2(t/\TH)^{1.30\pm0.02}$.  The barely significant
deviation of the exponent from $2\alpha=4/3$ may be due to the logarithmic
corrections expected in $d=2$ or the remnant effects of viscosity.

By way of comparison I have also done some corresponding simulations
on a binary liquid mixture.  This is achieved in regular DPD by
introducing two species of particle with an additional repulsion
between them \cite{DPDbin}.  The parameter sets in
Table~\ref{table-binary} were used, and the results are also shown in
Fig.~\ref{fig-scaling}.  The growth exponent is that expected for
diffusion-limited droplet coalescence in $d=2$ as has been seen in
previous work \cite{NC}.  Note the absence of a scaling collapse
between the two parameter sets in this case.

It may be worthwhile to interject a brief comment about the opposite
limit to the bubble coarsening regime, namely a dispersion of drops in
a vapour phase.  If the vapour is sufficiently dilute, coarsening can
procees by \emph{ballistic} drop coalescence, as has been reported
previously for a Lennard-Jones fluid \cite{LSR}.  Following Bray
\cite{Bray}, simple scaling arguments suggest a growth law
$\bigl(\phi t\sqrt{\kT/\rho}\bigr){}^{2/(d+2)}$.  I have confirmed the
validity of various aspects of this growth law in $d=2$ at $\phi=0.1$
and 0.05.

In summary therefore, I have presented arguments which show a
fundamental difference in phase coarsening between binary liquid
demixing and vapour-liquid phase separation.  Simulations of a soft
sphere fluid strongly support the idea of an inertial hydrodynamic
(IH) bubble coarsening regime for an off-critical vapour-liquid
quench. The relevance of the critique of Grant and Elder \cite{GR} is
not immediately apparent in this case, nevertheless there has to be a
mechanism to dissipate the interface energy.  It may be
that it is converted to kinetic energy in the fluid, and subsequently
lost in viscous dissipation by a turbulent cascade, as suggested for
the critical quench in a binary liquid mixture \cite{Kendon}.  But, in
real cavitation, a significant amount of the interface energy is
converted into pressure waves \cite{cavitation}.  Thus another
possibility is that fluid compressibility plays a role, and the
interface energy is (partly) dissipated through sound waves.  The
subtleties of these mechanisms do not show up in the statistics of the
bubble size distribution since the data in Fig.~\ref{fig-scaling}
indicates the mean bubble size follows the IH growth law very closely
indeed, but the signature may perhaps be sought in the statistics of
the flow fields.

I thank A. Louis, R. D. Groot and M. E. Cates for discussions.

\pagebreak

\onecolumn

\mediumtext

\begin{table}
\begin{tabular}{ccccccccc}
Set&$A$&$B$&$\rd$&$\rho$&$\eta$&$\sigma$%
   &$\LH=\eta^2/\rho\sigma$&$\TH=\eta^3/\rho\sigma^2$\\
\tableline
`4'&$-40$&$27$&$0.75$%
   &$3.954(1)$&$3.19(3)$&$2.83(7)$%
   &$9.09\times10^{-1}$&$1.03\times10^{0{}\phantom{-}}$\\
`5'&$-40$&$18$&$0.75$%
   &$4.950(1)$&$2.53(3)$&$5.29(8)$%
   &$2.44\times10^{-1}$&$1.17\times10^{-1}$\\
`6'&$-40$&$12$&$0.75$%
   &$6.333(1)$&$1.69(2)$&$9.44(6)$%
   &$4.78\times10^{-2}$&$8.55\times10^{-3}$\\
\end{tabular}
\caption{Parameter sets for many-body DPD, for vapour-liquid
coexistence, and properties of the liquid phase determined by
simulation (in all cases, the figure in brackets is an estimate of the
error in the final digit). The vapour phase density is vanishingly
small.  The final two columns show the hydrodynamic length and time
scales.  Units are DPD units, where $m=\rc=\kT=1$.
\label{table-many}}
\end{table}

\begin{table}
\begin{tabular}{ccccccccc}
Set&$A_{11}=A_{22}$&$A_{12}$&$\rho$&$\eta$&$\sigma$%
   &$\LH=\eta^2/\rho\sigma$&$\TH=\eta^3/\rho\sigma^2$\\
\tableline
`4a'&$\phantom{0}5$&$40$%
    &$4$&$1.15(5)$&$3.90(4)$%
    &$8.48\times10^{-2}$&$2.50\times10^{-2}$\\
`4b'&$20$&$50$%
    &$4$&$1.10(3)$&$6.55(4)$%
    &$4.62\times10^{-2}$&$7.76\times10^{-3}$\\
\end{tabular}
\caption{Parameter sets for ordinary DPD, for symmetric binary liquid
mixtures..\label{table-binary}}
\end{table}


\begin{references}
\bibitem{many-note}{Many-body DPD is described in terms of forces
here, but a Hamiltonian formulation is also possible which indicates
the link betweeen the weight function and the many-body force law
\cite{PF}.}
\end{references}
\end{document}